\documentclass[review]{elsarticle}

\usepackage{lineno,hyperref}
\modulolinenumbers[5]
\usepackage{multirow}%
\usepackage{booktabs} 
\usepackage{xcolor}%
\journal{Journal of \LaTeX\ Templates}

\begin{document}

\begin{frontmatter}



\title {Prediction of Brain Tumor Recurrence Location Based on Multi-modal Fusion and Nonlinear Correlation Learning}





\author[mymainaddress]{Tongxue Zhou}
\author[mysecondaddress]{Alexandra Noeuveglise}
\author[mysecondaddress]{Romain Modzelewski}
\author[third]{Fethi Ghazouani}
\author[mysecondaddress]{Sébastien Thureau}
\author[mysecondaddress]{Maxime Fontanilles}
\author[third]{Su Ruan\corref{mycorrespondingauthor}}
\cortext[mycorrespondingauthor]{Corresponding author}
\ead{su.ruan@univ-rouen.fr}

\address[mymainaddress]{School of Information Science and Technology, Hangzhou Normal University, Hangzhou 311121, China}
\address[mysecondaddress]{Department of Nuclear Medicine, Henri Becquerel Cancer Center, Rouen, 76038, France}
\address[third]{Université de Rouen Normandie, LITIS - QuantIF, Rouen 76183, France}

\begin{abstract}
Brain tumor is one of the leading causes of cancer death. The high-grade brain tumors are easier to recurrent even after standard treatment. Therefore, developing a method to predict brain tumor recurrence location plays an important role in the treatment planning and it can potentially prolong patient's survival time. There is still little work to deal with this issue. In this paper, we present a deep learning-based brain tumor recurrence location prediction network. Since the dataset is usually small, we propose to use transfer learning to improve the prediction. We first train a multi-modal brain tumor segmentation network on the public dataset BraTS 2021. Then, the pre-trained encoder is transferred to our private dataset for extracting the rich semantic features. Following that, a multi-scale multi-channel feature fusion model and a nonlinear correlation learning module are developed to learn the effective features. The correlation between multi-channel features is modeled by a nonlinear equation. To measure the similarity between the distributions of original features of one modality and the estimated correlated features of another modality, we propose to use Kullback-Leibler divergence. Based on this divergence, a correlation loss function is designed to maximize the similarity between the two feature distributions. Finally, two decoders are constructed to jointly segment the present brain tumor and predict its future tumor recurrence location. To the best of our knowledge, this is the first work that can segment the present tumor and at the same time predict future tumor recurrence location, making the treatment planning more efficient and precise. The experimental results demonstrated the effectiveness of our proposed method to predict the brain tumor recurrence location from the limited dataset. 
\end{abstract}


\end{frontmatter}

\section{Introduction}
\label{sec1}
Brain tumors are one of the most aggressive cancers in the world \cite{jiang2021novel}. Brain tumors consist of abnormally growing tissue resulting from the uncontrolled multiplication of cells \cite{miglani2021literature, nazir2021role}. Gliomas are the most common brain tumors that arise from glial cells. According to the WHO (World Health Organization) \cite{louis20212021}, primary brain tumors can be classified from grade I to grade IV. Grades I and II are considered low-grade glioma (LGG), while grades III and IV are high-grade glioma (HGG). Low-grade tumors are less aggressive, and they come with a life expectancy of several years. High-grade tumors are much more aggressive, and they have a median survival rate of less than two years \cite{lefkovits2022hgg, magadza2021deep}. According to the National Brain Tumor Foundation (NBTF), the number of people in developed countries who die from brain tumors has increased as much as 300\% \cite{abd2019review, el2014computer}. Automated and accurate segmentation of brain tumors is of vital importance for clinical diagnosis \cite{wang2021transbts}.

The optimal treatment for brain tumor is multimodal \cite{lapointe2018primary}. First, surgery is operated to remove the visible tumor regions, which is the most common treatment for brain tumors. And in most of the cases, surgery is the only necessary treatment, and the detailed surgical plan will depend on the tumor size and location. Then, the radiation therapy uses X-rays and other forms of light energy to destroy cancer cells in malignant tumors or to delay the progression of a benign brain tumor. And it is usually recommended after surgery and can be combined with chemotherapy \cite{kumar2021malignant}; In addition, targeted drug therapies are frequently paired with surgery or radiation to treat metastatic brain tumors. Finally, supportive care, such as psychological support, anti-edema and anti-epileptic treatments, is associated. Despite this standard treatment, brain tumors still recurrent \cite{parvez2014diagnosis,liu2019applications,mendes2018targeted}. Therefore, developing a method to predict the brain tumor recurrence location is of significance for anticipating the choice of appropriate treatment and following personalized treatments. 

Medical imaging techniques, such as Magnetic Resonance Imaging (MRI), CT scans, and Positron emission tomography (PET), among others, play a crucial role in the diagnosis of the tumors. These techniques are used to locate and assess the progression of the tumor before and after treatment \cite{magadza2021deep}. Magnetic Resonance Imaging (MRI) is considered the standard modality for diagnosis and prognosis of brain tumors due to its high resolution, soft tissue contrast, and non-invasive characteristics \cite{tiwari2020brain, mecheter2022deep, li2021whole}. Moreover, there are four commonly used MR sequences: T1-weighted (T1), contrast-enhanced T1-weighted (T1c), Fluid Attenuation Inversion Recovery (FLAIR) and T2-weighted (T2) images. In this work, we refer to these different sequences as modalities. Different MR modalities can highlight different sub-regions, for example, FLAIR and T2 can highlight the edema region, T1 and T1c can focus on the tumor core region. Therefore, multi-modalities can provide complementary information to visualize the brain tumor \cite{wadhwa2019review, icsin2016review, zhou2019review}. In addition, multiple time points of multi-modalities can provide essential information for locating brain tumor recurrence, and improving treatment planning. However, it's challenging to collect multiple time points images in clinical practice since it requires a long-term follow-up experiment. To this end, in this paper, we propose a novel deep neural network to predict brain tumor recurrence location in the pixel level using only two time point multi-modal images.

The rest of the paper is organized as follows. Section \ref{sec2} introduces the related works on the prediction of brain tumor recurrence location. Section \ref{sec3} presents our proposed method. Section \ref{sec4} introduces the experimental settings. Section \ref{sec5} presents the experimental results. The conclusion and future work are summarized in Section \ref{sec6}.

\section{Related Works}
\label{sec2}
In recent years, many research works have been proposed for brain tumor recurrence prediction. We generally classify them into two main groups: Mathematical Model (MD)-based method and Deep Learning (DL)-based methods. 

MD-based methods are mainly based on the Reaction-Diffusion (RD) model, which basically describes the tumor growth using two terms: tumor cells diffusion/invasion and tumor cells proliferation \cite{domschke2014mathematical,swan2018patient,liu2014patient,mi2014prediction}. Although these methods achieved good performance, they are applied to every subject independently without considering the population common features. In addition, these models usually have a few parameters that are insufficient to learn the complex behaviour of tumor growth process \cite{elazab2020gp}. Also, these methods require images of at least two time points and corresponding segmented tumors. 

Lately, deep learning, as an alternative of learning hierarchical features, has proven high performance in training models with millions of parameters. DL-based methods have also been widely used in the field of tumor growth prediction \cite{ezhov2023learn}. For example, Wang et al. \cite{wang2019toward} proposed a deep neural network to predict the spatial and temporal trajectories of lung tumor based on weekly MRI data. First, a convolutional neural network is presented to extract relevant tumor features. Then, a recurrence neural network is designed to analyze the trajectories of tumor evolution. Finally, an attention model is introduced to weight the importance of weekly observations and produce predictions. Zhang et al. \cite{zhang2017convolutional} proposed to use deep convolutional neural networks to predict the subsequent involvement regions of a tumor by directly representing and learning the two fundamental processes of tumor growth (cell-invasion and mass-effect). The proposed invasion network can learn the cell invasion from information related to metabolic rate, cell density and tumor boundary derived from multimodal imaging data. The proposed expansion network can model the mass-effect from the growing motion of tumor mass. Elazab et al. \cite{elazab2020gp} proposed a stacked 3D generative adversarial networks to predict the growth of gliomas. In addition, the segmented feature maps are used to guide the generator for better generated images. Petersen et al. \cite{petersen2019deep} proposed to use probabilistic segmentation and representation learning to predict brain tumor growth without specifying an explicit model. However, these methods mentioned above are designed to predict the brain tumor growth using longitudinal MRI. It usually requires at least three time point images. First, at least two time point images are used to learn the tumor growth model. Then, the model can be used to predict the tumor shape for the next time point. 

In this work, our objective is to predict the topography of recurrence after surgery, which varies from patient to patient, in order to anticipate personalized treatments. Only two time point multi-modal images are available in our task, the inputs are the initial time point multi-modal images in the stage of diagnosis, and the ground-truth is the tumor recurrence regions after surgery. It is very challenging to directly predict the tumor recurrence location using one time point multi-modal images. Nonetheless, it is important to physicians to know what will happen to the patient after the surgery. Therefore, we propose a multi-modal fusion module and a nonlinear correlation learning module to predict brain tumor recurrence location from limited dataset modalities. To the best of our knowledge, there are no available related works reported in the literature. Our preliminary conference version appeared at 26th International Conference on Pattern Recognition (ICPR 2022) \cite{zhou2022prediction}. This journal version is a substantial extension, including 1) A novel multi-scale and multi-channel feature fusion model is developed to recalibrate the feature representations along modality-attention and spatial-attention paths. 2) A larger training dataset is used to enhance the feature learning ability of the network. 3) This new work has improved the results of the previous works. Our main contributions can be described as follows:

(1) A novel multi-scale multi-channel feature fusion model is proposed to selectively extract useful feature representations from multi-modalities.

(2) The proposed nonlinear correlation learning module can model the multi-source correlation between multi-modalities in the latent space. And a KL divergence is designed to measure the similarity between the multi-modal feature distributions. 

(3) The proposed method is capable to jointly segment the present brain tumor before surgery and predict the future brain tumor recurrence location after surgery.

(4) The experimental results demonstrated the effectiveness of the proposed method, which opens the door to the prediction of brain tumor recurrence location in images.

\section{Proposed method}
\label{sec3}

\subsection{Motivation}
Traditional deep convolutional neural networks require a large amount of labeled data for training to achieve optimal performance. However, collecting a large quantity of medical data to study a specific medical issue is difficult. Transfer learning is to apply the knowledge learned from a related source task with large amounts of training data to a target task with comparatively insufficient training data in a certain way \cite{cui2020semantic,8877822}. Some low-level features, such as the edges and shapes of objects, are relevant and can be shared by transferring parameters. Collecting the multiple time points patient information for recurrence prediction is particularly challenging. To solve the problem of lack of data, we propose to use transfer learning to segment present brain tumor regions as well as to predict the brain tumor recurrence regions in the images. The pre-trained network can learn rich semantic features from larger dataset, then these features can be transferred to the present network trained with limited training data. In this way, the problem of limited training data can be alleviated, and the network can extract the meaningful features for following segmentation and prediction tasks.

\subsection{Network architecture}
\begin{figure*}[htb]
  \centering
  \centerline{\includegraphics[width=\textwidth]{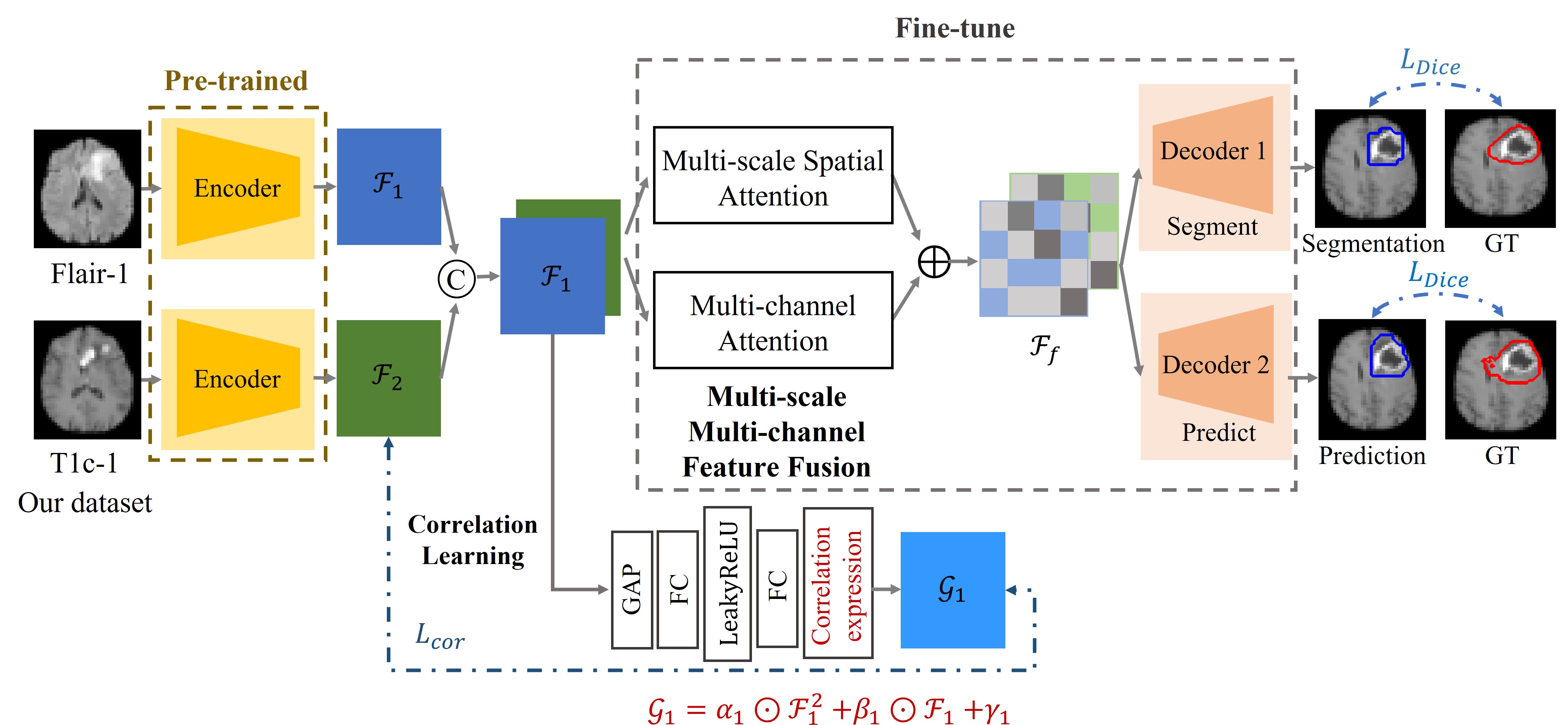}}
\caption{The overview of the proposed network architecture. The network includes two encoders for feature extraction, a multi-scale multi-channel feature fusion model, a nonlinear correlation learning module and two decoders. It is noted that BraTS 2021 dataset only has one time point image at diagnosis time. Our private dataset has two time point images, however, only the initial time point images (Flair-1 and T1c-1) are used for training. $\copyright$ denotes concatenation, $\oplus$ denotes the addition, $\odot$ denotes element-wise product.}
\label{fig1}
\end{figure*}

The overview of the proposed method is depicted in Figure \ref{fig1}. First, a multi-modal brain tumor segmentation network is trained on public BraTS 2021 dataset \cite{menze2014multimodal}, consisting of 1251 cases. Through which the network can extract rich semantic features. Then, the pre-trained encoders are transferred to the private dataset for feature extraction. Following that, the multi-scale multi-channel feature fusion model is introduced to learn the effective feature representations. In addition, a nonlinear correlation learning module is presented to discover the latent multi-source correlation between the input MR modalities as well as to extract the correlated features. Finally, two decoders are applied to simultaneously obtain the tumor segmentation before surgery and recurrence location prediction after surgery. The segmentation and prediction decoders have the same architecture, and the skip connections are used between them to improve the prediction result. It is noticed that both results are not at the same time point, and we focus on the recurrence location prediction result. 

\begin{figure*}[htb]
  \centering
  \centerline{\includegraphics[width=0.8\textwidth]{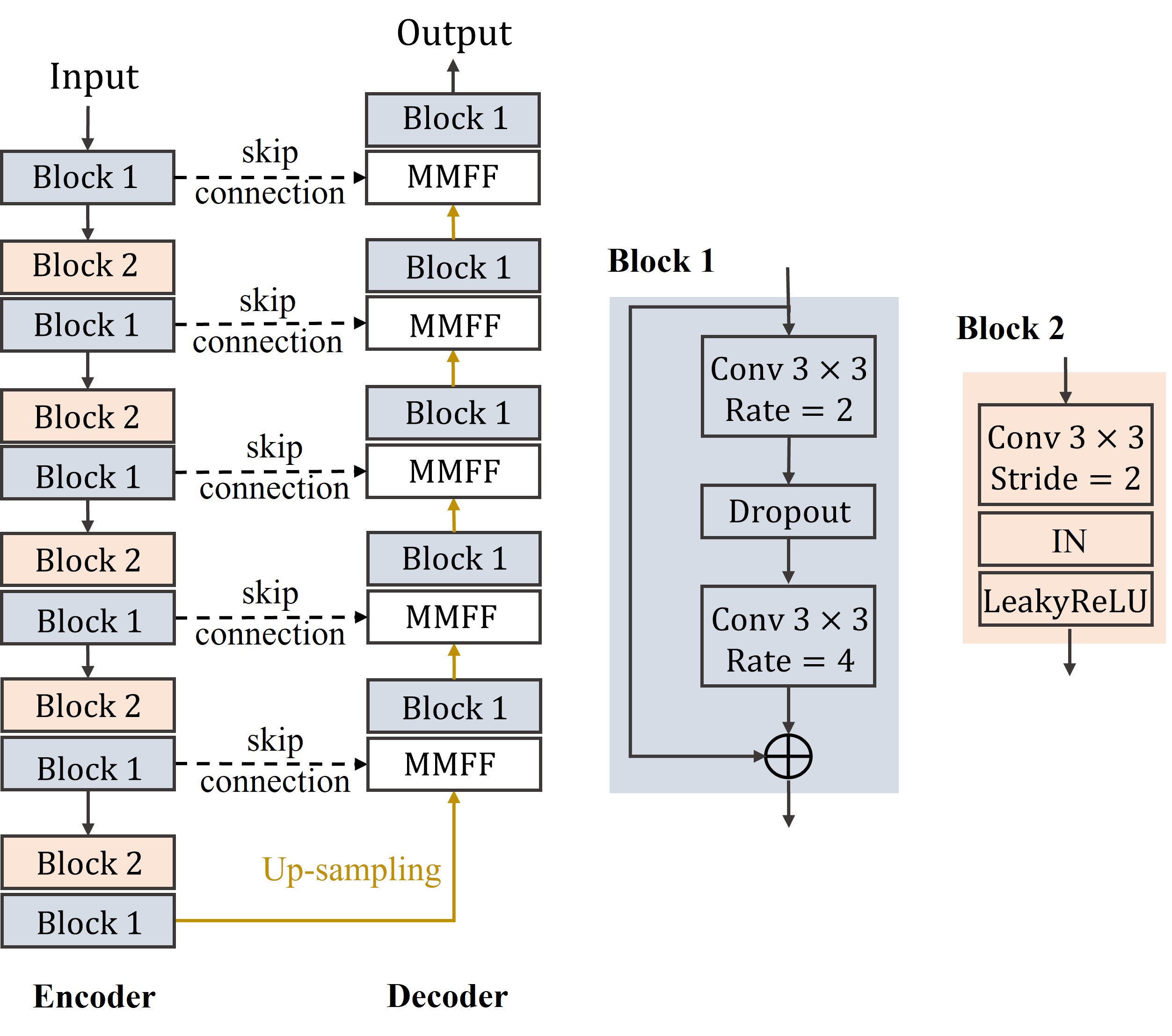}}
\caption{The architectures of the proposed encoder and decoder. IN refers to instance normalization. To obtain different receptive fields for segmenting different regions in the image, the dilated convolution is applied on both encoder and decoder to obtain multi-scale features. Besides, to maintain the spatial information, the convolution with stride=2 is applied to replace the pooling operation.}
\label{fig2}
\end{figure*}

The detailed architectures of the encoder and decoder are presented in Figure \ref{fig2}. The encoder consists of a group of dilated convolution layers to obtain different receptive fields for segmentation. In order to maintain the spatial information, the convolution with stride=2 is applied to replace the pooling operation. The decoder includes an up-sampling layer, a multi-scale multi-channel feature fusion model, and a group of dilated convolution layers. The skip connections are used to combine the features of the encoder and decoder.  

\begin{figure*}[htb]
  \centering
  \centerline{\includegraphics[width=\textwidth]{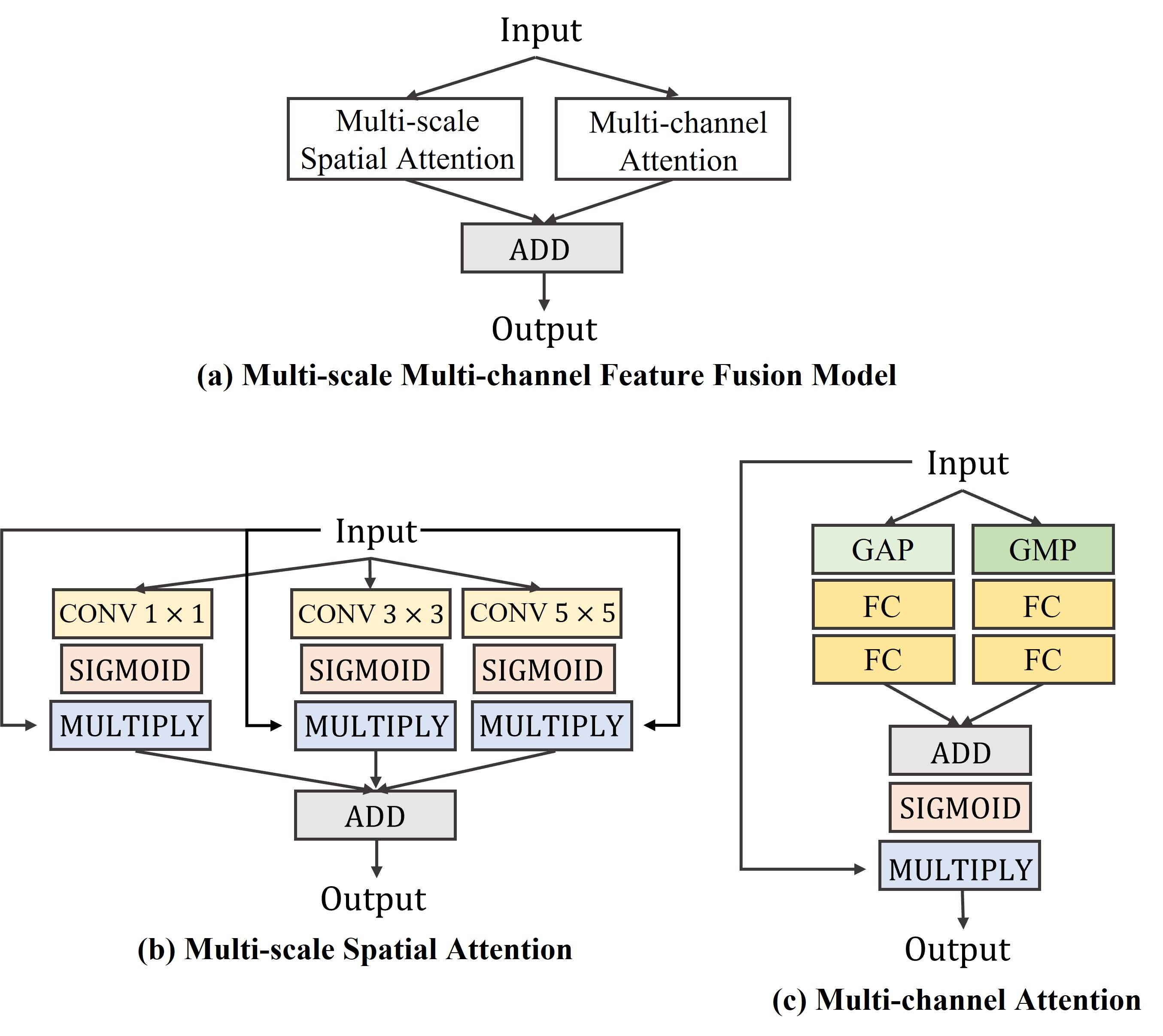}}
\caption{The overview of the proposed (a) Multi-scale Multi-channel Feature Fusion (MMFF) model, consisting of (b) multi-scale spatial attention module and (c)  multi-channel attention module.}
\label{fig3}
\end{figure*}

\subsection{Multi-scale Multi-channel Feature Fusion Model}
Recently, the attention model is widely used in the field of image classification \cite{chen2021crossvit, aghalari2021brain}, semantic segmentation \cite{huang2019ccnet} and object detection \cite{dai2021dynamic}. For multi-modal brain MR images, different MR modalities can focus on different tissue structures and underlying anatomy. Therefore, designing an effective feature fusion method to learn the important features from different modalities is essential for brain tumor segmentation and prediction. In this work, we propose a novel multi-modal fusion model, named Multi-scale Multi-channel Feature Fusion (MMFF) model. Since different scale features can provide different receptive fields, which can eventually improve the segmentation performance. Therefore, we ameliorate the previous single-scale spatial attention module \cite{zhou2022prediction} to a multi-scale spatial attention module. Besides, to improve the previous single-channel attention module, we extend it to a dual-channel attention module. The architecture of MMFF model is depicted in Figure \ref{fig3}. 

In the multi-scale spatial attention module, the input feature is first fed to three different convolution layers to focus on different receptive fields. Then the sigmoid function is employed to obtain the different spatial attention weights, indicating the importance of spatial position in the feature maps. By multiplying the input feature with the spatial attention weights, the multi-scale spatial feature can be obtained. In the multi-channel attention module, a global average pooling and a global max pooling are utilized to extract the global average and max feature information from the input feature. By using addition and sigmoid function, the channel attention weights can be obtained, indicating the importance of modality. The multi-channel attention feature can be obtained by multiplying the attention weights with the input feature. The two weighted features are finally added to achieve the fused feature. In this way, the network can selectively recalibrate features along multiple spatial and channel paths, and extract useful feature representations from multi-modalities. 


\subsection{Correlation learning}
Since the same tumor regions can be observed in the two MR modalities, there exists a statistical correlation between modalities, which has been proved in our previous work \cite{9412796}. In the previous work, we assume it is a linear correlation. In this work, we improve it to a nonlinear correlation, which is defined in Equation \ref{eq1}. The comparison results between the two correlation expressions are presented in Section \ref{5.4}. In the correlation learning module, a global average pooling followed by two fully connected layers is first used to learn the correlation weights $\{\alpha_i, \beta_i, \gamma_i\}$. Based on these weights and the input feature $\mathcal{F}_i$, the estimated correlated feature $\mathcal{G}_i$ can be obtained. To constrain the original feature distribution of modality $j$ and the correlated feature distribution of modality $i$ to be close as possible, we proposed a Kullback–Leibler divergence-based loss function (see Equation \ref{eq2}). The comparison results between different divergences are presented in Section \ref{5.3}. In this way, the network can learn the correlated features for the following segmentation and prediction.

\begin{equation}
    \mathcal{G}_i = \alpha_i \odot {\mathcal{F}_i}^2+\beta_i \odot  \mathcal{F}_i+ \gamma_i 
\label{eq1}
\end{equation}

\noindent where $\mathcal{F}_i$ and $\mathcal{G}_i$ are the original feature and estimated correlated feature of modality $i$, $\alpha_i$, $\beta_i$ and $\gamma_i$ are the correlation weights to be learnt, $\odot$ denotes element-wise product.

\begin{equation}
	L_{cor} =P(\mathcal{F}_j) log\frac{P(\mathcal{F}_j)}{Q(\mathcal{G}_i)}
	\label{eq2}
\end{equation}

\noindent where $P(\mathcal{F}_j)$ and $Q(\mathcal{G}_i)$ are probability distributions of the original feature of modality $j$ and estimated correlated feature of modality $i$, respectively. Lower $L_{cor}$, lower the divergence.

\subsection{Loss function}

Dice loss (see Equation \ref{eq3}) is used for tumor recurrence location prediction. For the tumor segmentation, the loss function is defined in Equation \ref{eq4}.

\begin{equation}
    \ L_{Dice}=1-2\frac{\sum_{i=1}^N\ p_{i} g_{i}+\epsilon} {\sum_{i=1}^N (p_{i} + g_{i})+\epsilon}
\label{eq3}
\end{equation}

\noindent where $N$ is the set of all examples, $p_{i}$ is the prediction result, $g_{i}$ is the ground-truth, and $\epsilon$ is a small constant to avoid dividing by 0.

\begin{equation}
    \ L_{seg}= L_{Dice}+ \phi L_{cor}
\label{eq4}
\end{equation}
\noindent where $\phi$ is the trade-off parameter, which is empirically set as 0.1 in our work.

\section{Experimental settings}
\label{sec4}
\subsection{Datasets}
\label{sec4.1}
We collected a clinical MRI dataset of 51 cases from our collaborative hospital (Henri Becquerel Center\footnote{https://www.becquerel.fr/}, France). The dataset contains 2 time points images. The first time point images are the initial MR images before surgery, the second one is the tumor recurrence images after surgery. Each time point has two MR modalities: FLAIR and T1c. The provided data have been pre-processed by doctors: co-registered to the same anatomical template, interpolated to the same resolution ($1 mm^3$) and skull-stripped. The size of the images ranges between $512\times512\times110$ to $512\times512\times211$. The ground truth including enhancing and necrosis tumor regions is annotated by doctors. In addition, we complete our data by another dataset of 16 cases from BraTS 2015\footnote{https://sites.google.com/site/braintumorsegmentation/home/brats2015}. Each case has 2 time points images, each time point has four modalities: FLAIR, T1, T1c and T2, and four ground truth labels: enhancing tumor, non-enhancing tumor, necrosis and edema. The size of the images is $240\times240\times155$. To be consistent with our private dataset, we use two modalities: FLAIR and T1c. And unify both enhancing and necrosis as one ground-truth label. All 67 cases are followed by the same pre-processing. Since the limited dataset, we used 2D images and resize them to $128\times128$. Then the N4ITK \cite{avants2009advanced} method is used to do the bias field correction for MRI data. Intensity normalization is applied to normalize each modality to a zero-mean, unit-variance space. 

\subsection{Implementation details}
\label{4.2}
The proposed network is implemented using Keras with one Nvidia Tesla V100. The model is trained using Adam optimizer. The initial learning rate is 0.0005, it will reduce with a factor 0.5 with a patience of 10 epochs for the pre-trained model and the target task. To avoid over-fitting, early stopping is used if the validation loss is not improved over 50 epochs for the pre-trained model and the target task. We randomly split the dataset into 80\% training, and 20\% for testing. 

\begin{table}[htb]
\centering
\caption{Segmentation and prediction performance of three different methods in terms of DSC, HD and Sensitivity.}
\label{tab1}
\vspace{0.2cm}
\resizebox{\textwidth}{!}{%
\begin{tabular}{ccccc}
\hline
Methods & Task & DSC (\%) & HD (mm) & Sensitivity (\%)\\ \hline

 & Segmentation & 54.1&9.0 & 56.3\\
\multirow{-2}{*}{(1): Direct} & Prediction & 47.4 & 9.8 &43.4\\ \hline

& Segmentation & \textbf{64.2} & \textbf{5.3} &57.8\\ 
\multirow{-2}{*}{(2): Test on pre-trained model} & Prediction & 52.2 & \textbf{7.0} &44.9\\ \hline
 
 & Segmentation & 63.2 & 7.9&\textbf{74.6}\\ 
\multirow{-2}{*}{(3): Transfer learning (Ours)} & Prediction &\textbf{63.5} & 7.6 &\textbf{69.2}\\ \hline
\end{tabular}}
\end{table}

\begin{table}[htb]
\centering
\caption{Ablation experiment results among different methods in terms of DSC, HD and Sensitivity}.
\label{tab2}
\vspace{-0.1cm}
\resizebox{\textwidth}{!}{%
\begin{tabular}{ccccc}
\hline
Methods  & Task & DSC (\%) & HD (mm) &Sensitivity (\%)\\ \hline
\multirow{2}{*}{Baseline} & Segmentation & 61.2 &\textbf{5.6} &61.3 \\ & Prediction &49.3&7.9 &44.1 \\ \hline
\multirow{2}{*}{Baseline + MMFF}  & Segmentation &60.9  & 6.2 &56.7\\ & Prediction & 50.8&7.9&43.2  \\ \hline
\multirow{2}{*}{Baseline + MMFF + Correlation learning} & Segmentation & \textbf{63.2} &7.9 &\textbf{74.6} \\ & Prediction & \textbf{63.5} & \textbf{7.6} &\textbf{69.2}\\ \hline
\end{tabular}}
\end{table}
\begin{figure*}[htb]
  \centering
  \centerline{\includegraphics[width=1.0\textwidth]{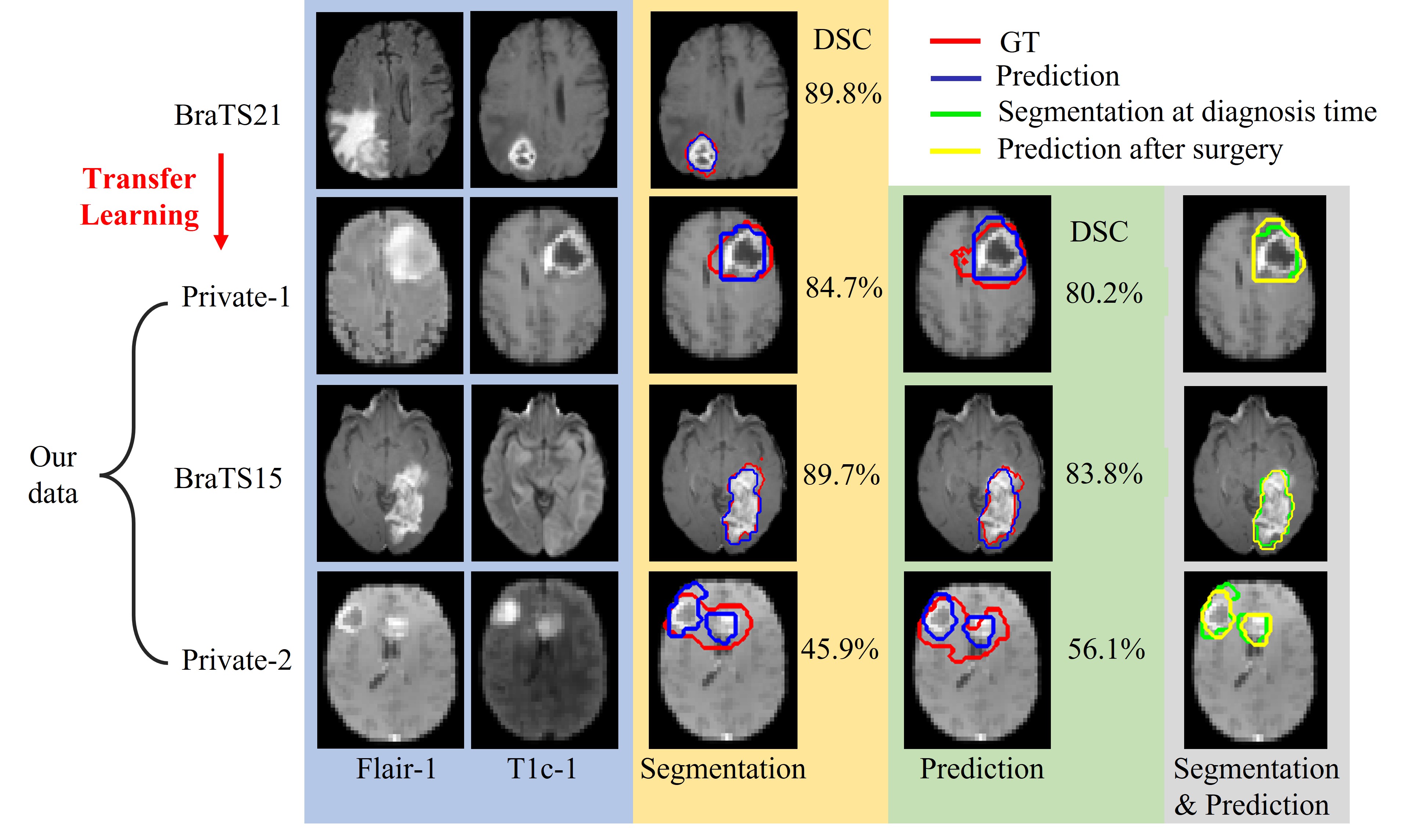}}
\caption{Visualization of the segmentation and prediction performance of our proposed method with three patients: two from our private dataset, and one from the BraTS 2015 dataset. Flair-1 and T1c-1 denote the modalities at the first time point, which are used for training our model. The last column jointly presents our segmentation and prediction results.}
\label{fig4}
\end{figure*}

\subsection{Evaluation metrics}
\label{sec4.3}
To evaluate our method, we used three commonly used evaluation metrics: Dice Similarity Coefficient (DSC), Hausdorff Distance (HD) and Sensitivity as used in the BraTS challenge. The better result will have a larger DSC, a smaller HD, and a larger Sensitivity.

\begin{equation}
 DSC= \frac {2TP}{2TP+FP+FN}
\end{equation}

\noindent where $TP$ represents the number of true positive voxels, $FP$ the number of false positive voxels, and $FN$ the number of false negative voxels. 

\begin{equation}
		HD = \max\{\max\limits_{s\in S}\min\limits_{r\in R} d(s,r), \max\limits_{r\in R}\min\limits_{s\in S} d(r,s)\}
\end{equation}
\noindent where $S$ and $R$ are the two sets of the surface points of the prediction and the real annotation, and $d$ is the Euclidean distance.

\begin{equation}
Sensitivity=\frac{TP}{TP+FN} 
\end{equation}

\section{Experimental Results}
\label{sec5}
\subsection{Quantitative analysis}
\label{sec5.1}
We first conduct the experiments to see the effectiveness of transfer learning in terms of Dice Similarity Coefficient (DSC), Hausdorff Distance (HD) and Sensitivity. The results are presented in Table \ref{tab1}. We can observe that directly using the limited dataset (only private data), noted ``Direct'' in Table \ref{tab1}, to do the segmentation and prediction can't achieve satisfactory results, especially on Hausdorff Distance. In addition, if we use the pre-trained network to test our dataset without the transfer learning, for the segmentation task, we can achieve 64.2\% in terms of DSC, 5.3 in terms of HD and 57.8\% in terms of Sensitivity; for the prediction task, we can obtain 52.2\% in terms of DSC, 7.0 in terms of HD and 44.9\% in terms of Sensitivity. With the assistance of transfer learning, the accuracy to segment brain tumors can obtain 74.6\% in terms of Sensitivity, with an improvement of 29.1\% in terms of Sensitivity, compared to the method (2). The accuracy to predict the recurrence location can achieve 63.5\% in terms of DSC, and 69.2\% in terms of Sensitivity, with an improvement of 21.6\% in terms of DSC and 54.1\% in terms of Sensitivity, compared to the method (2).

We also conduct the ablation experiments to demonstrate the effectiveness of the proposed strategies, including multi-scale multi-channel fusion model and nonlinear correlation learning, the comparison results are presented in Table \ref{tab2}. It can be observed that the proposed components can help to improve the baseline method with an improvement of 3.3\% in terms of DSC and 21.7\% in terms of Sensitivity for segmentation and 28.8\% in terms of DSC, 3.8\% in terms of HD, and 56.9\% in terms of Sensitivity for prediction. The experiments demonstrate the importance of the proposed strategies.

\subsection{Qualitative analysis}
\label{sec5.2}
We select several examples to visualize the segmentation and prediction results in Figure \ref{fig4}. The DSC values are shown in each example. The first row presents the segmentation results of the pre-trained model. The last three rows present the segmentation and prediction results of our transferred model on our private dataset and BraTS 2015 dataset. To have a good visualization of our results, we jointly present the segmentation and prediction results in the last column. From the first example in the second row, we can obtain a good segmentation result (84.7\% in terms of DSC), and a good prediction result (80.2\% in terms of DSC). From the second example in the third row, it can be observed that the proposed method can obtain good segmentation (89.7\% in terms of DSC) and good prediction results (83.8\% in terms of DSC). For the third example in the last row, the segmentation result is not really good (45.9\% in terms of DSC), and the recurrent location is not well predicted (56.1\% in terms of DSC) neither. We explain that from the two input modalities, two bright tumor regions are easily to be observed, while the low contrast tumor regions are not easy to be observed. Thus, our trained model only focuses on these two bright tumor regions, and ignore the other regions. The reason is that the feature learning ability of the network is not strong enough, in the future work, we will enhance the feature learning ability of our network to further improve the results. In conclusion, the above experimental results prove that using transfer learning can indeed help to jointly segment brain tumor and predict the tumor recurrence location.

\subsection{Analysis on the divergence functions}
\label{5.3}
To measure the similarity of the feature distributions, we compare three commonly used divergences: Kullback–Leibler (KL) divergence, Jeffreys divergence (Equation \ref{eq11}) and squared Hellinger divergence (Equation \ref{eq12}). The compared results are shown in Table \ref{tab3}. It can be seen that KL divergence achieves the best segmentaion and prediction performance. In particular, KL divergence outperforms Jeffreys divergence by 4.5\% and 37.1\% in terms of DSC for segmentation and prediction, respectively, 11.6\% in terms of HD for prediction, respectively, and 29.1\% and 75.6\% in terms of Sensitivity for segmentation and prediction, respectively. Jeffreys divergence and squared Hellinger can achieve better HD for segmentation task than Kullback Leibler, however, the other results are unsatisfactory.

\begin{equation}
D_J(P \| Q )= (P(\mathcal{F}_j)-Q(\mathcal{G}_i))(\ln{P(\mathcal{F}_j)}-\ln{Q}(\mathcal{G}_i))
\label{eq11}
\end{equation}

\begin{equation}
H^2(P, Q) =  2(\sqrt{P(\mathcal{F}_j)}-\sqrt{Q(\mathcal{G}_i)})^2
\label{eq12}
\end{equation}

\noindent where $P(\mathcal{F}_j)$ and $Q(\mathcal{G}_i)$ are the original feature distributions of modality of $j$ and correlated feature distributions of modality $i$, respectively.

\begin{table}[htb]
\caption{Comparison results among different divergence functions in terms of DSC, HD and Sensitivity.}
\vspace{0.2cm}
\label{tab3}
\resizebox{\textwidth}{!}{%

\begin{tabular}{cccccc}
\hline
Divergence functions               & Task         & DSC (\%) & HD (mm) &Sensitivity (\%)\\ \hline
\multirow{2}{*}{Kullback–Leibler}  & Segmentation &\textbf{63.2}&7.9&\textbf{74.6} \\
                                   & Prediction   &\textbf{63.5}&\textbf{7.6}& \textbf{69.2} \\ \hline
\multirow{2}{*}{Jeffreys}          & Segmentation &60.5&\textbf{6.5}&57.8        \\
                                   & Prediction   & 46.3    &8.6& 39.4       \\ \hline
\multirow{2}{*}{squared Hellinger} & Segmentation &61.8      &\textbf{6.5}&60.6      \\
                                   & Prediction   &47.2      &7.8 &39.2     \\ \hline
\end{tabular}}
\end{table}

\subsection{Analysis on the correlation expressions}
\label{5.4}
For the correlation learning module, we compare different correlation expressions including a linear correlation expression (Equation \ref{eq-linear}) and a nonlinear one (Equation \ref{eq1}). From Table \ref{tab4}, we can observe that the nonlinear correlation expression can achieve better segmentation accuracy, especially for tumor recurrence prediction, with an improvement of 8.4\% in terms of DSC, and 19.7\% in terms of Sensitivity. We explain that the nonlinear correlation expression can better describe the feature distribution. 

\begin{equation}
    \mathcal{G}_i = \alpha_i \odot  \mathcal{F}_i+ \gamma_i 
\label{eq-linear}
\end{equation}

\noindent where $\mathcal{F}_i$ and $\mathcal{G}_i$ are the original feature and estimated correlated feature of modality $i$, $\alpha_i$ and $\gamma_i$ are the correlation weights.

\begin{table}[htb]
\caption{Comparison results among different correlation expressions in terms of DSC, HD and Sensitivity. }
\label{tab4}
\vspace{0.2cm}
\resizebox{\textwidth}{!}{%
\begin{tabular}{ccccc}
\hline
Correlation expressions                     & Task         & DSC (\%) & HD (mm) &Sensitivity (\%) \\ \hline
\multirow{2}{*}{Linear}     & Segmentation & 58.5     & \textbf{7.1} &55.2     \\
                            & Prediction   &58.6     & \textbf{7.4 }  &57.8    \\ \hline
\multirow{2}{*}{Non-linear} & Segmentation &\textbf{63.2}&7.9 & \textbf{74.6}     \\
                            & Prediction &\textbf{63.5}&7.6&  \textbf{69.2}     \\ \hline
\end{tabular}}
\end{table}
\section{Conclusion}
\label{sec6}
In this paper, we propose a novel method to predict the brain tumor recurrence location in multi-MR modalities. To address the limited dataset, we propose to utilize transfer learning. Through the pre-trained network on the larger training dataset, the  present network can extract rich semantic features for segmentation and prediction. To fuse the multi-modalities, a multi-scale multi-channel feature fusion model is proposed. The proposed model can selectively recalibrate features along both spatial and channel paths to benefit the segmentation. Moreover, to learn the multi-modal correlation, a nonlinear correlation learning module is proposed. Besides, a Kullback–Leibler divergence-based loss function is introduced to measure the similarity between the distributions of the extracted features from two modalities. Minimizing the loss function is to maximize the similarity between the two distributions. To this end, the network can learn the correlated features for the following segmentation and prediction. On the one hand, the proposed network can segment the present brain tumor before surgery; on the other hand, it can predict the location of brain tumor recurrence in the future after the patient undergoes surgery. The experimental results demonstrate the proposed components can help to improve the baseline method with an improvement of 3.3\% in terms of DSC and 21.7\% in terms of Sensitivity for segmentation and 28.8\% in terms of DSC, 3.8\% in terms of HD, and 56.9\% in terms of Sensitivity for prediction. In addition, the proposed method opens the door for tumor recurrence location prediction from the small dataset. It can help clinical doctors diagnose brain tumor and therapists design personalized treatment planning for future tumor recurrence. 

Our method has some limitations that inspire future directions. It does not yet give perfect results, however, it opens promising perspectives to develop personalized treatment planning in a more relevant way by jointly considering the spatial prediction of recurrence, and the segmentation of the tumor at the diagnostic stage. Other recent feature fusion methods, e.g. self-attention model, can be investigated to benefit the work.

In the future, we would like to generate artificial recurrent MR images to improve the prediction of the brain tumor recurrence location in images. In addition, we would like to take the clinical factors and genomic information into account to ameliorate the prediction accuracy. And we will collect much larger datasets to validate and further improve our method.

\section*{Acknowledgment}
This work was financed by National Natural Science Foundation of China (No. 62206084) and European project FEDER PREGLIO (No. 21E03175).

\bibliography{mybibfile}

\end{document}